\documentclass{article}
\usepackage[utf8]{inputenc}
\usepackage{amsfonts}
\usepackage{amsmath}
\usepackage{graphicx}

\title{Tutorial on the development of AI models for medical image analysis}
\author{Thijs Kooi\\
Lunit}
\date{}

\begin{document}
\maketitle

% \begin{abstract}
%   Many systems
%   
%   Most research
%   
%   This tur
% \end{abstract}

\section{Introduction}
\label{sec:introduction}
The idea of using computers to read medical scans was introduced as early as 1966 \cite{lodwick1966computer}. However, limits to machine learning technology meant progress was slow and stalled after studies revealed merit may be limited \cite{fenton2007influence}. The Alexnet breakthrough in 2012 sparked new interest in the topic and in 2017 already 100s of papers on the use of deep neural networks for medical image analysis were published \cite{litjens2017survey}. Startups in the artificial intelligence space have also increased drastically with now over 100 products in the European market \cite{van2021artificial}. Additionally, some AI systems have been shown to operate at or above human level and are allowed to work autonomously \cite{nagendran2020artificial, abramoff2018pivotal, salim2020external}. \\

In spite of success for a few diseases and modalities, many challenges remain. Medical data are diverse, scarce and rife with long tail samples. New imaging techniques are constantly being developed, meaning AI solutions have to be newly developed or adapted to new domains. Research typically focuses on the development of specific applications or techniques, clinical evaluation, or meta analysis of clinical studies or techniques through surveys or challenges. However, limited attention has been given to the development process of improving real world performance. In this tutorial, we address the latter and discuss some techniques to conduct the development process in order to make this as efficient as possible. \\

\section{Model development}
\label{sec:model_development}
\begin{figure}
 \centering
 \includegraphics[width=0.7\textwidth]{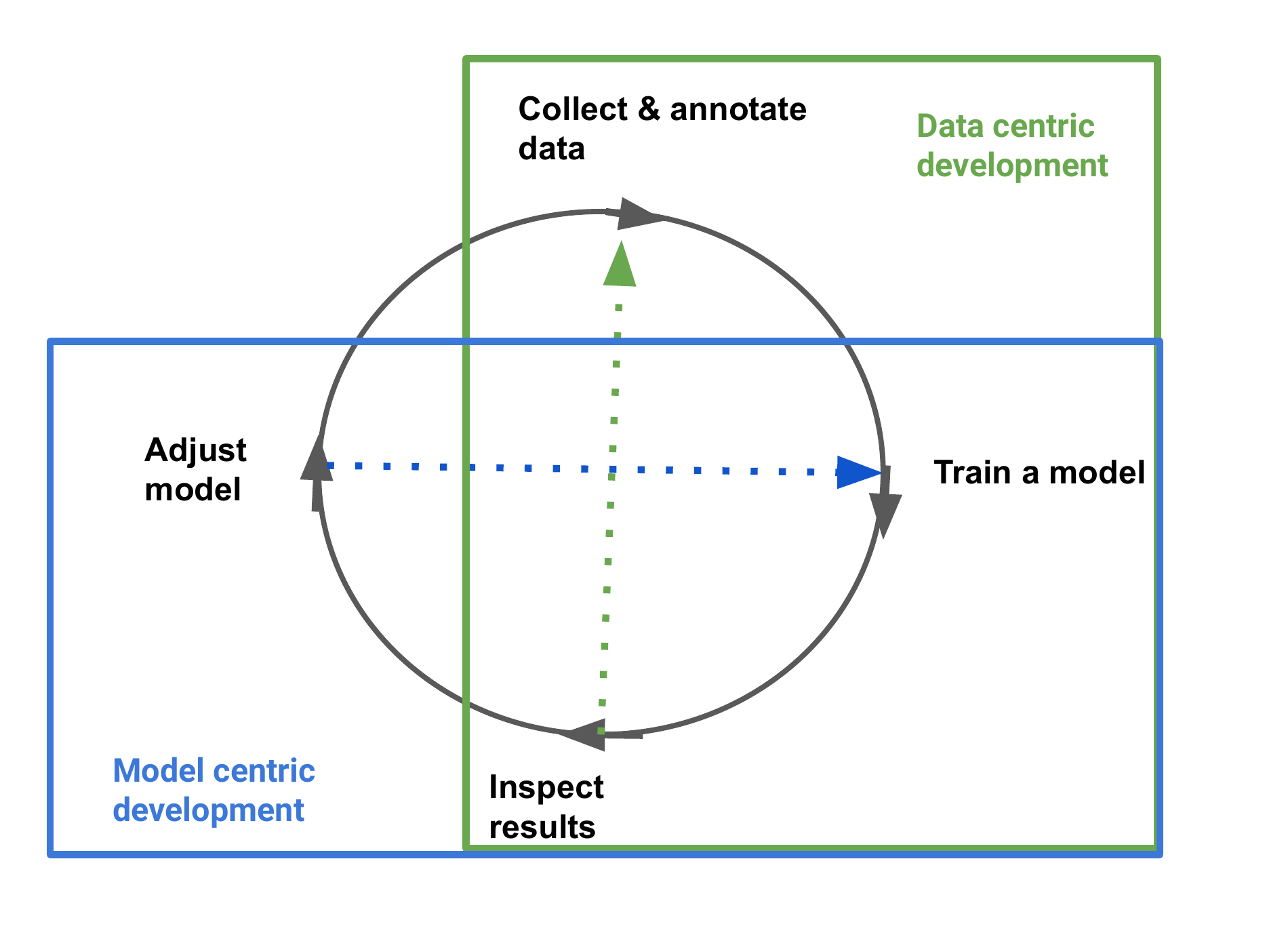}
 \caption{Illustration of a typical model development cycle. When a clinically relevant problem has been identified, we start by collecting data and optionally annotating it. After that, we can train our first model on the data, evaluate it on a validation set and determine our next steps, which can be either changing the data, the model architecture or the training process.}
 \label{fig:model_development_cycle}
\end{figure}

After a clinically relevant problem has been identified, development typically starts with data collection and (optionally) annotation. We can then split the data into a training, validation and test set and train a first model. During development, the model is evaluated on the validation set, if the model is good enough, we can test it and report the final performance. If not, we have to find some way to improve the model. The whole process can be seen as a while-loop that we execute until some convergence criterion is met, illustrated in figure \ref{fig:model_development_cycle}. 

The model development process can roughly be divided into two approaches, which are almost always combined in practice:
\begin{enumerate}
 \item {\bf A model centric approach}, illustrated with the blue box in figure \ref{fig:model_development_cycle}. This is the 'typical' model development process as it is done in many academic settings and challenges. In the extreme case, it assumes a static dataset and evaluation metric and iteratively changes the model architecture/training procedure until results are satisfactory. This roughly comprises the following changes:
 \begin{enumerate}
    \item Finding a good fit to the data for a specific architecture. Are the weights and biases well initialized? Do gradient updates of the weights look reasonable? Do we increase or decrease the learning rate?, etc. 
    \item Finding a good architecture. This can include, but is not limited to: adding loss functions that force the model to better fit the data, adding inductive biases (like making the model invariant to certain types of texture, geometrical transformation), making the model sensitive to location, sensitive to context like different views of the same body part, etc. 
  \end{enumerate}
  \item {\bf A data centric approach}, illustrated with the green box in figure \ref{fig:model_development_cycle}. This approach puts more focus on the data and less on the architecture. In the extreme case, the model architecture is fixed and we only change what data is presented and how the data is presented to the model. \\\
  This can include collecting more data from a specific type where the model makes frequent errors \cite{cohn1996active}, annotating more findings of a specific type, correcting annotations \cite{bernhardt2022active}, augmenting the training data with artificial samples, training on a subset of the data that we think better represent the test distribution, re-ordering the way in which samples are presented \cite{bengio2009curriculum}, fixing issues in preprocessing, etc. This type of AI engineering was less popular until a few years ago, but became more dominant recently, especially with increased involvement from industry, where resources are typically more suitable for handling large scale data.
\end{enumerate}

As there are many excellent tutorials on model centric development, especially on the fitting part \cite{karpathy2019recipe}, this tutorial will have a strong focus on the latter. In practice both are needed for a well working solution and the two are hard to separate. To get the most out of our data, we need a good model and to get a good model we need good data. For now, we will assume that we have some training, validation and test data and an initial model that has been evaluated on our validation data. We want to determine a good next step.

\section{Model diagnosis}
\label{sec:model_diagnosis}
\subsection{Model fitting}
A good first step to see how the model can be improved would be to check if it actually fit well to the data. If the model does not fit well, many of the more sophisticated analysis techniques will not be useful. A simple cooking recipe for fitting a model would be:
\begin{itemize}
 \item Overfit on a small subset (e.g., 5 images) of the data. Make sure we can get a training error of 0. This means the model should be relatively bug free and is at least able to learn a simple problem. Since there are many excellent high level deep learning packages nowadays \cite{paszke2019pytorch}, the implementation part should be fairly straightforward. 
 \item Apply the same, but then to a larger chunk of the training set. If we can (almost) overfit the entire training set, this means the model has enough capacity and a suitable architecture to model the problem
 \item Add regularization until we see the train and validation performance are almost at the same level. 
\end{itemize}
If step 1 fails, we need to open up the model and see what is happening during training. The first thing to do is to tweak the learning rate (e.g., in the form of a grid search). Visualizing the weights and gradients using tools like W\&B, Tensorboard or MLflow is also often useful, as this can tell us all the weights change and are distributed well accross the network. Other approaches would be to simply use automatic hyperparameter tuning. For more elaborate tutorials on this topic please refer to e.g., \cite{karpathy2019recipe}. \\

When the model fits well to the data, we can do more extensive analysis and find pain points of the current data and architecture. 

\subsection{Machine learning's basic assumption}
Most machine learning algorithms used these days follow the assumptions that the data are independently, identically distributed. This means the data points should be independent from each other and train and test data comes from the same distribution. In practice, the latter assumption almost never holds, meaning the model can underperform in unseen data. Models for medical image analysis are no exception and can fail to generalize to variables like different manufacturers, patient populations and disease subtypes. \\

\begin{figure}
 \centering
 \includegraphics[width=0.7\textwidth]{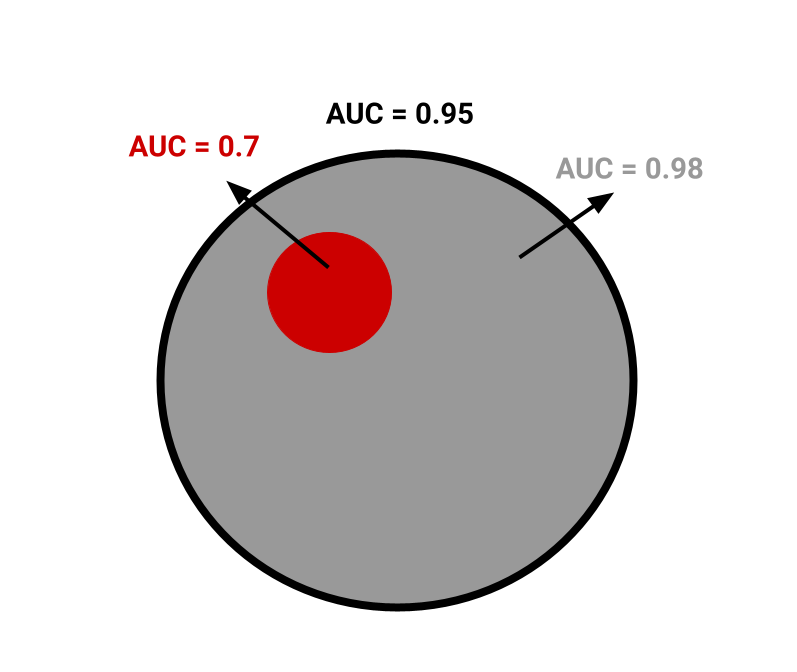}
 \caption{A good place to start when diagnosing a model is to look for generalization issues. In practice, it is often possible to isolate a subset of the data where the model does not work well on. If we can improve that, we improve the overall performance. In this example, the area under the ROC (AUC) on the entire validation set is 0.95, however on a small subset of the data the performance is only 0.7. If we can improve that, we can improve the general performance and ultimately real-world performance.}
 \label{fig::02_subgroup_venn_diagram}
\end{figure}

If the model is fit well to the data, a good place to look for ways to improve the model is by looking for generalization issues, which are not revealed when looking only at aggregate statistics over an entire validation set. To get a sense of where the model fails to generalize, we can assume that we can split our data into separate groups and that we can find a group on which the performance is worse than the rest of the data. If we improve this group, the aggregate performance is expected to increase. This is illustrated in figure \ref{fig::02_subgroup_venn_diagram}. To better understand this concept, let's consider a simple example. \\

\begin{center}
\noindent\fbox{%
    \parbox{0.9\textwidth}{%
        Assume that we have a dataset with images taken from devices from two different manufacturers, let’s call them manufacturer A and B. Now assume that 90\% of the data in our evaluation set is of manufacturer A and 10\% from manufacturer B. Let's assume we are interested in the sensitivity of our model for detecting the specific disease. If we get a sensitivity of 100\% on manufacturer A and a sensitivity of 50\% on manufacturer B, we still get a sensitivity of 95\% on average.\\
        
        Looking only at the aggregate would not have revealed we perform poorly on manufacturer B. If we would deploy this to a site which has a distribution $[10, 90]$ for manufacturer A and B respectively, we would get a sensitivity of 90 * 50 + 10 * 90/ 100 = 54\%, which is pretty bad. This phenomenon is also referred to as spectrum bias in the medical literature: the performance of a medical test, treatment or something similar can vary widely if the distribution of the data changes \cite{ransohoff1978problems}. \\
        
        If we can improve the performance of our model on manufacturer B, the general performance is expected to go up, especially in populations where manufacturer B is common. 
    }
}
\end{center}

\subsection{Attributes and annotations}
The manufacturer used above is just an example of a data attribute that typically comes with medical dataset \footnote{and not only medical datasets, natural image datasets might have images recorded with different camera’s, different lighting conditions etc \cite{russakovsky2015imagenet}, though this is often not registered and collected in datasets}. This type of data can be seen as additional annotations that we can exploit. We often have metadata in the DICOM header or from additional reports about recording conditions (such as the pixel spacing of the detector, the peak kilovoltage, the number of slices in the scan, etc), some characteristics of the patient (like the age, BMI or ananmesis), information about disease subtype, etc. Lastly, we can perform feature transformations of the data ourselves, for example by computing image features like sharpness, contrast, orientation, etc.\\

\subsection{Marginalization and subgroups}
Let's look at this example more mathematically. Assume we have three random variables: the disease we are trying to classify $C$, the output of our model $Y$ and the manufacturer $M$ (one of the attributes we can use to split the data). The sensitivity of a model is the probability that the model outputs the value 1, given the class has value 1, or $P(Y = 1 | C = 1)$. Implicitly, this is a marginal distribution over many different subgroups in the data, that we can generate from attributes we have. For example, for manufacturers:
\begin{equation}
 P(Y = 1 | C = 1) = \sum^{M}_{m = 1}P(Y = 1 | C = 1, M = m) 
\end{equation}
A key observation here is that {\it every performance metric we report is always a marginal over many variables}, either observed (i.e., known unknowns) or not observed (unknown unknowns). If we assume multiple attributes (such as the patient age $A$, the image brightness $B$, etc), the marginal becomes 
\[
 P(Y = 1 | C = 1, A = a, B = b, \ldots)
\]

As we saw in the example above, changing the distribution of the data we evaluate our model on can make a big difference in aggregate performance if there is a clear difference in performance between different populations. We therefore need to make sure the dataset is sampled properly and is a representative sample from the target population and always make sure to look into generalization issues for relevant variables. One further complication is that the variables are often correlated, which could be the result of confounding.

\subsection{Confounding}
Confounding is a common phenomenon in any type of data analysis and refers to a variable that affects both an intervention, for example a treatment like chemotherapy or exposure to some toxic substance like cigarette smoke and an outcome, such as the survival rate or the development of lung cancer. This is also best explained with a simple example: \\

\begin{figure}
 \centering
 \includegraphics[width=0.35\textwidth]{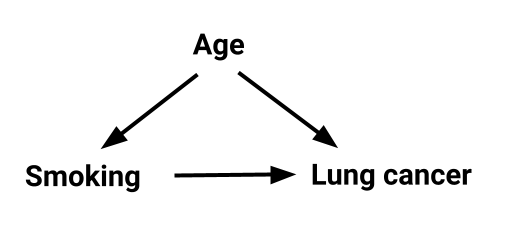}
 \includegraphics[width=0.55\textwidth]{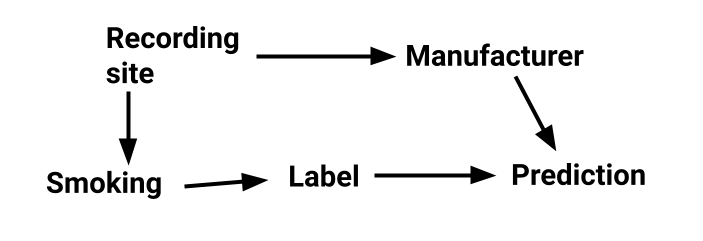}
 \label{fig:confounding_example}
 \caption{(Left) A textbook example of confounding. Age increases the risk for lung cancer and smoking is more common in older people. When we observe the increased lung cancer morbidity in smokers, we need to normalize for age, before we can conclude this is due to smoking. \\
 (Right) An example of confounding in a typical medical image analysis problem. In this scenario, the manufacturer of the machine that is used affects the performance of the model and is correlated with the specific disease type (label). One way this could happen would be through a 'recording site' variable that affects both the label and the device manufacturer that is used, for example because specific disease types and manufacturers are more common in certain sites.}
\end{figure}

\begin{center}
\noindent\fbox{%
    \parbox{0.9\textwidth}{%
        Assume we want to detect lung nodules in chest CT scans. We have data from two sites: A and B. We know that site A only used Siemens scanners and in site B only Phillips scanners are used. We also know that the incidence of smoking is higher near site A and therefore specific types of nodules associated with smoking are common, let's refer to them as type 1 nodules. 
        
        We evaluate our model on a test set from both sites and observe our model is worse on data from site A. In this case, this can have three reasons (1) the model is bad at detecting nodule's of type 1, (2) the model performs worse in scans taken with a Siemens scanner or (3) both. In this example the site is referred to as a {\it confounder}. This example is illustrated graphically in figure \ref{fig:confounding_example}. 
    }
}
\end{center}

A straightforward way to remove confounding during evaluation is to decorrelate the variables. In the example above, this would be trying to collect a dataset where the each scanner type has an equal number of type 1 and type 2 nodules. Another solution would be to split the data into four groups instead of two, conditioning on both the nodule and scanner type. In practice, this is not always possible unfortunately, because of limited data. Additionally, finding some shortcuts may require of domain knowledge, which may not be readily available to developers.

\subsection{Shortcuts}
A similar problem to confounding can occur during training, which can result in a specific type of generalization issues. Learning can be seen as looking for correlations between the input data and the variable we want to predict. This can also mean ‘spurious correlations’ (correlations that exist in the training but not in the test data) are learned, which have been referred to as ‘shortcuts’ \cite{geirhos2020shortcut} in recent literature, but are a well-known phenomenon \cite{jo2017measuring}. Again, this is best illustrated with an example. \\

\begin{center}
\fbox{%
    \parbox{0.9\textwidth}{%
        Imagine we have a dataset with two different manufacturers: manufacturer A and manufacturer B. Both of these look very different and can easily be discriminated if we would train a model to do that specifically (for example based on average brightness of the image). Now imagine we want to train a model to discriminate cases containing signs of some disease from normal cases. If it is far easier to discriminate between the manufacturers than the signs of disease, the model will learn a shortcut and could get a perfect classification performance on the training set. \\
        
        If we now take the same model and apply it to a test set with the same diseases and manufacturers, but now with a 50/50 split of manufacturers for each disease, that is, all samples with disease 1 have 50\% manufacturer A and 50\% manufacturer B, and likewise for disease 2, we will get an accuracy of 50\%. 
    }
}
\end{center}

Identifying and coming up with a solution for shortcuts is fairly straightforward {\it if} we have annotations for the features that are exploited. For example, if analysis reveals that in our training data 80\% of the chest radiographs with a pneumothorax contain a pneumothorax tube, the model may learn to detect the tube instead of the pneumothorax. We can decrease this correlation by ensuring the model also sees many normal chest radiographs with tubes and remove the shortcut. \\

In general however, we do not have annotations for all features in the image that the model could exploit, which means finding and solving shortcuts is much more difficult in practice \cite{ribeiro2016should, viviano2019saliency, degrave2020ai, beery2018recognition, adebayo2021post}. Shortcuts are sometimes revealed when the model is applied to out of distribution data, where the particular shortcut does not exist (for example, not all cases with a disease are taken with a specific manufacturer). Another way to reveal them would be through manual inspection of the data. 

\subsection{Manual inspection}
The methods defined above provide some guidelines to look for generalization issues. If we do not have annotations or if any transformations we make do not reveal any issues, the best option is typically to talk to a clinical collaborator if they can find some time to go through cases manually. They often have a good intuition about what can be improved. A simple recipe for this would look like:

\begin{itemize}
 \item Choose an operating point for the model, classify all samples in the validation set into positives and negatives (assuming a binary problem).
 \item Collect sets of {\it false positives} and {\it false negatives}
 \item For each of these sets, look for patterns in the data by going through them manually. 
\end{itemize}
These patterns could be for example: most of the false positives are generated in scans from patients with a specific body shape for which we have no annotation, perhaps our model is not invariant to this particular transformation, most of the false negatives seem to be in a subtype of cancer, the model generates false positives in cysts, abnormalities in the breast that resemble masses, the model seems to detect a tag used by a specific site, etc. Note that this type of analysis is essentially the opposite of looking for generalization issues by splitting the data into groups. We do not look for groups where the model performs worse, but for poorly classified samples and try to group them. \\

The steps above are some rough guidelines for model diagnosis, but unfortunately they do not always reveal useful points of improvement. In many cases, we simply do not have enough data. Medical data is difficult to acquire and most of it is set aside for training. Even if we can find some patterns, the subset where the model makes a mistake may be too small to draw conclusions. A simple solution would be to collect more data for validation, but this is not always possible in practice.  

\section{Solutions}
In this section, we provide two detailed real world examples of identified problems and solutions, followed by a few shorter examples. Just like the diagnostic process, the solutions presented can roughly be split into model and data centric solutions. The exact solution strongly depends on the problem. In some situations, a change in architecture may be the only thing that makes sense. For example, if the model does not have enough capacity or the patch size is too small to cover certain abnormalities, collecting more data will not get us anywhere. In other situations, for example when the model is simply poor at detecting a very specific subtype of a disease, the only solution may be the collect more data of that type. 

\subsection{Example from the detection of breast cancer in digital tomosynthesis}
The purpose of this project is to detect tumors in digital breast tomosynthesis scans. Apart from the class label (cancer yes/no), we have the following annotations:
\begin{itemize}
 \item Device manufacturer $\in \{ \text{Hologic}, \text{Siemens}, \text{GE} \}$
 \item Lesion size (binned into discrete groups)
 \item Radiological classification $\in \{ \text{soft tissue lesion}, \text{mass}\}$
 \item Breast density $\in \{ \text{A}, \text{B}, \text{C}, \text{D} \}$
\end{itemize}
The performance of the model is split in different subgroups and we evaluate the model on each group. We observe that the performance is comparable on all subgroups (e.g., Hologic vs. GE, soft tissue lesions vs. masses), but see a clear difference in lesion size: the model is much worse at detecting small lesions. There does not seem to be a strong correlation between variables, so we can conclude no confounders exist. \\

The model we work with is a segmentation model, and we see it is worse on small lesions because they receive smaller gradient updates during training, since the gradient correlates to the size in the image. We change the loss function to add more weight inversely depending on the size of the lesion and see the performance of the model increase for small lesions and the performance as a whole \footnote{Credit: Hyunjae Lee for coming up with the solution}. 

\subsection{Example from the detection of nodules in CT}
The purpose of this project is to detect nodules, which can be a symptom of lung cancer, in low dose CT scans. Apart from the class label, we have the following annotations:
\begin{itemize}
 \item Nodule size $\in \{ \text{small}, \text{medium}, \text{large} \}$
 \item Nodule type $\in \{ \text{Solid}, \text{subsolid}, \text{GGO} \}$
 \item Device manufacturer $\in \{ \text{Toshiba}, \text{Siemens}, \text{Philips}, \text{GE} \}$
 \item Slice thickness 
\end{itemize}

We split the internal validation set into four sets (corresponding to the four manufacturers) and compute performance metrics individually and obtain scores of $0.9, 0.8, 0.6$ and $0.7$ respectively and subsequently into different nodule sizes, for which we obtain scores of $0.9, 0.8, 0.6$, respectively. Looking at these scores, we can not know if the lack of performance is because of the manufacturer or because of the nodule size. It could simply be that people who are recorded using a Toshiba machine have larger nodules on average. It could be that another variable, for instance, the location where the scans are recorded is affecting both the manufacturer and the population (a confounder). \\

When generating 3D images from individual slices, the manufacturer applies a reconstruction kernel. Different reconstruction kernels are used by different vendors and additionally, the radiologist selects different kernels based on the abnormality they are trying to view. The kernel controls to some extent the sharpness of the image. To remove confounding from sites or manufacturers, we perform a feature transformation of the images by computing the sharpness directly, using a measure called acutance \footnote{Credit: Sanguk Park for coming up with the feature transform}. By looking at the performance split by different sharpness values, we see the model does not generalize to this variable and add unsharp mask and blurring augmentation to improve performance. 

\subsection{Further examples}
Model centric solutions
\begin{itemize}
%  \item We train a model on a large dataset of digital pathology scans, no matter what hyperparameters we tune, we see the train and validation AUC plateau at 0.54. We increase the model capacity.
 \item We find our model is worse at detecting a specific type of calcifications (very small signs of early stage breast cancer) in mammograms. We increase the resolution of the input images.
 \item We find our model is worse at detecting large lesions in chest CT scans. We increase the receptive field by adding dilated convolutions to our faster R-CNN architecture.
 \item We find our model does not generalize to a specific manufacturer, we add a module to our neural net that makes the model invariant to specific texture patterns.
\end{itemize}
Data centric solutions: 
\begin{itemize}
 \item We find our model does not generalize to specific manufacturer, we augment the training data by simulating the manufacturer where it does not generalize to.
 \item We find our model does not detect invasive lobular carcinoma well, we collect more data from this specific subtype of breast tumor.
 \item We find our model learns a shortcut between pneumothorax and pneumothorax tubes. We augment the data with fake pneumothorax samples.  
 \item We find our model's performance plateau's at an AUC of 0.7 for a specific finding in chest radiographs. We reaonnotate the training and validation data to remove noise in the annotation.
\end{itemize}

\section{Convergence}
\label{sec:convergence}
As mentioned in the introduction, the development process depicted in figure \ref{fig:model_development_cycle} is essentially a 'while' loop. These loops typically have a condition that determines when to stop. For the development process, we also need such rule and determine when our solution has converged to a performance that is sufficient. In a typical research application of a machine learning system, measuring convergence is relatively straightforward, by looking at loss curves. However, for the entire development process this is much more difficult. \\

\begin{figure}
  \centering
  \includegraphics[width=0.5\textwidth]{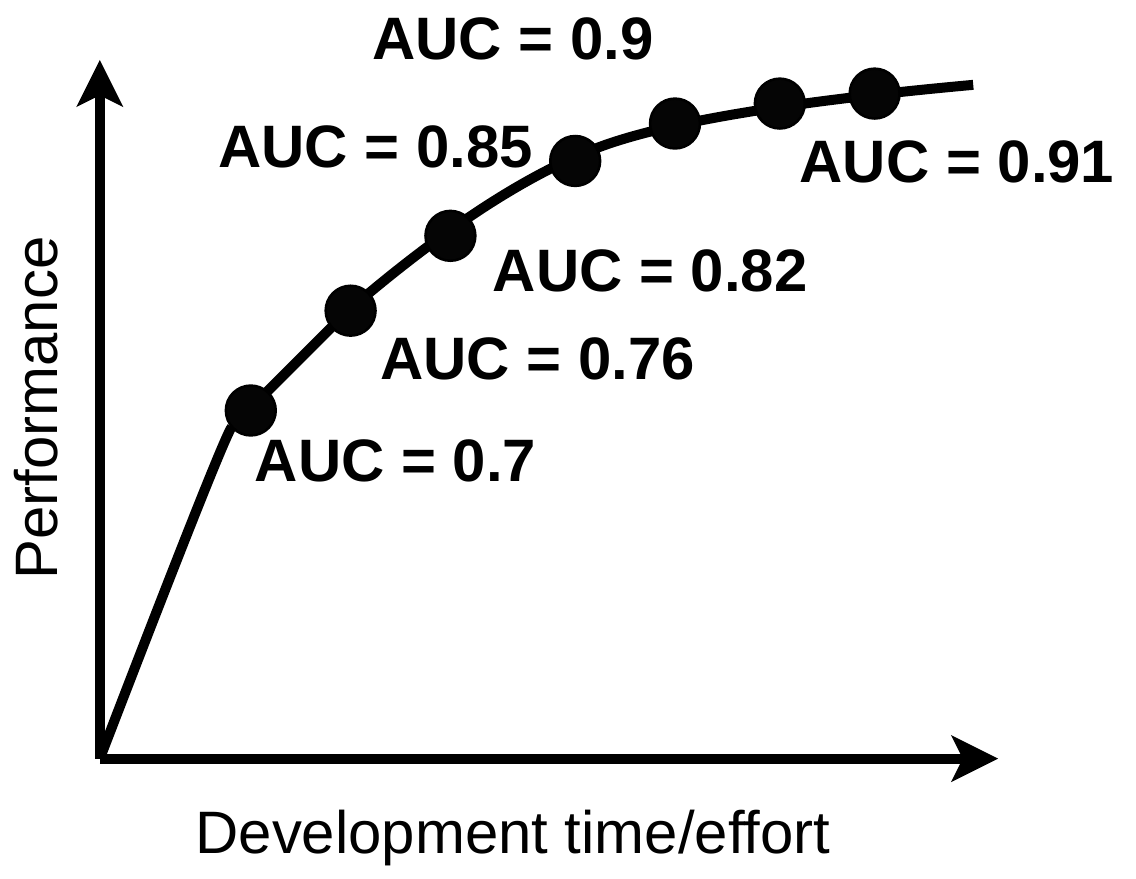}
  \caption{A (very) abstract curve of the typical model development process. On the Y-axis the performance of the model on some validation set and on the X-axis the development time. Each point represents a model update, which can include changes like better hyperparameters, a better model architectures or the addition of data.} 
  \label{fig:model_convergence}
\end{figure}

It is generally agreed upon that performance of some machine learning system asymptotically converges to some value as shown in figure \ref{fig:model_convergence} (although the curve is much less smooth in practice), however there is no guarantee that we can ever reach an AUC of 1 for every problem. Even, if this is possible, if this is truly asymptotic, we will see diminishing returns on R\&D investment: going from a AUC of 0.93 to 0.94 might take as much time as going from an AUC of 0.8 to 0.9. Secondly, continuing development can also mean overfitting the validation set if no other validation set re-use policies are in place \cite{gossmann2021test}. \\

Having a clear mathematical definition and theoretical derivation of when a model really converged for a specific problem is beyond the scope of this tutorial. Instead, we will discuss some general pointers related to (1) estimating where we are in the development process and (2) determining where we can and should get. To answer (1), a good first step would be to try to plot a learning curve, as is depicted in figure \ref{fig:model_convergence}. The curve will give us some idea of where we are, but it will not tell us much about how good the model can get. To get more information on that, it can be good to look into the types of labels that we have. 

\subsection{Label types}
In section \ref{sec:model_development} we already briefly discussed annotations, but did not go into detail on the specific type of labels. We can discriminate between two types of labels:
\begin{itemize}
 \item {\bf Labels based on human readers}. For example, annotations of cells in digital pathology slides, annotations of a pneumothorax in a chest radiograph, annotations of breast density in a mammogram, etc. For these labels, we have some inherent (sometimes referred to as aleatoric uncertainty in machine learning literature) \cite{kendall2017uncertainties} uncertainty, which results in label noise. 
 \item {\bf Labels based on some external test} that conveys more information than the particular scan. For example, a biopsy for tumors, a microbial culture for the detection of symptoms of bacterial diseases or a different scan that is known to be more accurate (e.g., an MRI or an FDG-PET scan). 
\end{itemize}
When using labels of the first type, it is unlikely the model will ever get an AUC of 1.0 unless the problem is so trivial that everyone will agree on the label. In this case, the performance is upper bounded by the variance within and between readers. If the model is within this variation, for example if the model gets an AUC of 0.94 when compared to labels of reader 1 and reader 2 gets an AUC of 0.93 when compared to the labels of reader 1, this may be a good time to stop. \\

When using labels of the second type, it is also unlikely we will ever be able to get an AUC of 1.0, again unless there are exceptional circumstances. Firstly, because an image is typically a many-to-one mapping from some real world object to lower dimensional space. It is theoretically possible that two abnormalities with different labels (e.g., a benign and a malignant nodule) are mapped to the same location in image space, meaning we can never tell the difference. Secondly, there are abnormalities that are picked up by external tests which are not visible to the human eye (e.g., occult breast cancers), but may be visible to a machine. For these abnormalities it is difficult to know if we can ever detect them using software. Human reader performance for the specific problem may give a good lower bound of how good our model can get, but the upper bound is typically unknown. \\

All above assumes we have fixed validation data. If we increase the size of our validation set, it is very well possible the model's performance decreases again on the new combined set, depending on how the data is sampled. How well we can measure converge also depends on the size of the validation set. If we only have 10 samples, the confidence bounds around results will be huge and we can probably not conclude much. Lastly, the convergence criterion should also depend on when the software would be useful in clinical practice. Even if we think model performance reached a plateau, we may not want to stop development if we think the model is still not good enough to be useful. Also in this case, looking at reader performance may help, as human reader performance is typically still the state-of-the art for many problems. 

\section{Summary}
In this tutorial, we gave a brief overview of a typical model development process and suggested some guidelines on how to perform that. Of course, these are general guidelines, other approaches may work and these guidelines may not work for all problems. We encourage other researchers and developers to share their development process. 

\section*{Acknowledgement}
Thanks to Sergio and Hesham for proofreading and providing suggestions.

\bibliographystyle{plain}
\bibliography{bibliography}

\end{document}